# A Neurally Constrained Computational Model of Context-Dependent Fear Extinction Recall and Relapse

Shreya K. Rajagopal*[1], Thad A. Polk[1]

*University of Michigan, Ann Arbor*

**Abstract:**

Exposure therapy, a standard treatment for anxiety disorders, relies on fear extinction. However, extinction recall is often limited to the spatial and temporal context in which extinction is learned, leading to fear relapse in new settings or after delays. Animal studies offer insights into fear extinction in humans. Computational models that integrate these findings into a neurally grounded framework, while generating testable hypotheses for humans, can bridge this gap. Current models either focus on neuron-level activity, limiting their scope, or abstract away entirely from neural mechanisms. They also often overlook the distinct contributions of cue and context in fear extinction and recall.

To address these gaps, we present ConFER, a neurally constrained model of fear extinction, recall, and relapse. ConFER integrates findings from the neural fear circuit, modeling distinct pathways for cue and context processing. These pathways independently activate positive and/or negative memory engrams in the basolateral amygdala, competing to determine the fear response. ConFER simulates fear renewal and spontaneous recovery across context combinations, while generating novel, testable predictions. Notably, it predicts counterconditioning may better prevent relapse than extinction in new contexts or after delays. By mechanistically modeling fear relapse, ConFER offers insights to improve exposure therapy outcomes.

**Introduction:**

When a neutral stimulus, such as a blinking light, is repeatedly paired with a fear-evoking unconditioned stimulus (US), such as a footshock, animals develop a conditioned fear response to the previously neutral stimulus—a process known as fear acquisition. When the conditioned stimulus (CS) is then repeatedly presented without the US, the fear response fades in a process called extinction. Exposure therapy, a widely used treatment for anxiety disorders, relies on fear extinction, but often yields mixed results[1,2], likely because extinction recall is highly context-dependent. In particular, extinction recall is often restricted to the specific spatial and temporal context where extinction occurred, causing fear to return in different contexts or after a long delay in the same context. Extensive neuroscientific research on fear extinction in animals offers valuable insights into the mechanisms underlying the relapse of fear post-therapy, but translating these findings to humans remains challenging. Neuroscientifically constrained computational models of fear conditioning phenomena that integrate key findings from the animal literature could help bridge this gap. Such models could not only replicate well-understood fear-conditioning phenomena but also provide novel hypotheses about the mechanisms underlying contextual extinction recall, which could inform clinical interventions.

Existing models of fear conditioning tend to adopt either a bottom-up or a top-down approach. A bottom-up approach models fear at the neuronal level, resulting in highly focused models that simulate the neuronal mechanisms underlying a specific set of experiments at the individual neuron level[3–5]. A top-down approach, on the other hand, reproduces a larger range of behavioral phenomena but at the cost of abstracting away from the details of the underlying neuroscientific mechanisms[6–9]. While each of these classes of models serves distinct and meaningful objectives, a middle-ground approach that leverages the advantages of both approaches holds promise for a unique set of insights.

Crucially, many models also overlook the role of context in fear extinction and its subsequent relapse. For bottom-up models, incorporating context processing might involve adding a simulation of hippocampal processing since the hippocampus is crucially involved in encoding and processing contexts[10]. However, owing to the complexity of simulating individual neuron activity in these models, they are typically limited to a single brain region such as the lateral amygdala, which encodes fear associations[3,4]. Even when multiple regions are included, the hippocampus is often excluded to avoid increasing complexity[5]. Top-down models, on the other hand, have more abstract learning rules that tend to treat extinction as the weakening of the CS-US association without separating cue and context information. For example, deterministic models like the Rescorla-Wagner and temporal difference models fail to account for the context-specific nature of extinction and treat it as context-independent unlearning[8,9]. In contrast, some probabilistic models of associative learning, like the latent cause theory model, approach extinction learning as a context-dependent phenomenon by attributing distinct latent causes to fear acquisition and extinction[6,7].

To address these gaps, we introduce the Context-Dependent Fear Extinction Recall Model (ConFER)—a computational framework that integrates known neural circuitry to capture key behavioral phenomena in fear processing, including context-dependent fear relapse. The model architecture, summarized in Figure 1, includes distinct pathways for processing cue and context inputs, which drive competitive interactions between positive and negative neuronal populations in the basolateral amygdala (BLA)[11]. Importantly, the BLA module incorporates reward-responsive neurons that form extinction memory engrams[12], aligning with recent evidence that extinction learning recruits the brain's reward circuitry. By incorporating differential learning rates for cue and context associations, ConFER can account for the fact that emotional associations change across time and context combinations, capturing key behavioral effects like fear renewal and spontaneous recovery.

In this paper, we provide a comprehensive description of the ConFER model's architecture and mechanisms, detailing how it simulates key context-dependent return of fear phenomena. We demonstrate that it can simulate fear acquisition and extinction learning across multiple experiment trials. We then simulate two key return-of-fear effects: fear renewal, where fear returns in contexts other than the extinction context, and spontaneous recovery, where fear re-emerges after a delay in the extinction context. We compare the predictions derived from the simulations to empirical findings from animal experiments. Additionally, we compare the effectiveness of fear extinction (CS presented without the US) to counterconditioning (CS presented paired with a positive reward) across various context combinations in order to derive a novel empirically testable prediction. Finally, we discuss ConFER's contributions, limitations, and future research directions.

Note that in this paper, we use the term "engram" to refer to a set of specific neurons that are activated during memory recall and can generate the memory if stimulated directly, aligning with the concept of engram cells as popularized by Tonegawa and colleagues[13,14]. This differs from the original usage of the term "engram", introduced by Semon and later popularized by Lashley, which refers to a hypothetical change in neural tissue to account for the persistence of memory[15].

**Model Overview:**

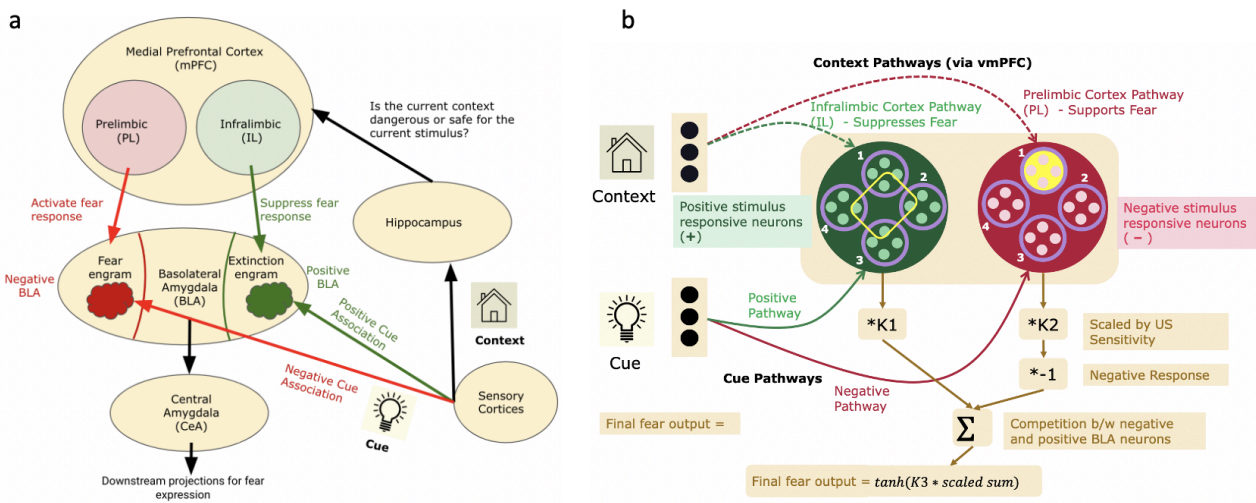

**Figure 1:** (a) A simplified schematic of the circuit of fear in the brain. Cue information travels directly to the basolateral amygdala (BLA), activating either the fear or extinction engram based on cue associations. Context information travels to the hippocampus and then to the BLA via the Infralimbic Cortex to suppress fear or the Prelimbic cortex to activate fear. There is competition between the fear and extinction engrams in the BLA, following which the central amygdala (CeA) has downstream projections to initiate the appropriate behavioral response (b) Computational model architecture. Context and cue inputs are connected to stored engrams in the positive (green) and negative (red) stimulus-responsive BLA populations. Purple circles indicate engrams stored in each BLA population, with the numbers corresponding to engram indices. The filled yellow engram indicates activation because the corresponding US (Shock) is present during an experimental trial. The yellow square indicates a selection of a random subset of BLA-positive neurons to form an extinction engram. The scaled sum of activations of each BLA population is used to determine the final fear level, which is computed as a tanh function.

The ConFER model presents a neurally constrained framework to propose biologically plausible mechanisms for the context-dependent return of fear after extinction, a critical challenge in exposure therapy for anxiety disorders. Building on previous computational approaches, ConFER explicitly distinguishes between cue and context processing pathways, emphasizing that cue associations form rapidly and remain stable over time, while context associations are slower to form and decay more easily.

A key assumption of the model is that extinction engrams form within reward-responsive neurons in the positive BLA population, highlighting the overlap between extinction and reward circuitry. ConFER bases this assumption on the findings of a recent study by Zhang et. al. that demonstrates that extinction engrams are formed and stored within the reward-responsive neuronal population in the positive BLA, with the neurons recruited to form the extinction engram being functionally interchangeable with reward-responsive neurons[12]. Zhang et. al. use their findings to propose that the omission of an expected aversive stimulus is rewarding[12]. In ConFER, extinction engrams for each cue form by recruiting a random set of neurons from the positive BLA, which are activated via their associations with the extinction context.

ConFER simulates both fear renewal (fear returning in non-extinction contexts)[16] and spontaneous recovery (fear returning in the extinction context after a delay)[17]. Additionally, the model proposes that counterconditioning—pairing a fear-conditioned cue with a positive unconditioned stimulus (US)—can lead to a lower return of fear compared to extinction, reducing both spontaneous recovery and renewal effects.

**O1. Core Components and Inputs**

The model processes two distinct inputs on each experimental trial: a *cue input* and a *context input*.

- **Cue Input**: Represents sensory information about a conditioned stimulus, such as a light or tone. This input is analogous to sensory cortices and is depicted by a lightbulb icon in Figure 1b.

- **Context Input**: Represents features of the environmental context in which the cue is presented. This input is conceptually linked to the hippocampal CA1 region, which encodes context information and is depicted by a house icon in Figure 1b.

Both inputs project to the basolateral amygdala (BLA), where emotional associations are encoded. The BLA consists of two distinct neuronal populations that respond to different types of stimuli:

- **Positive BLA Population**: This population contains neurons that respond to positive stimuli and encode safety or reward-related associations. In Figure 1b, the positive BLA population is indicated by the green circle, with existing unconditioned stimulus (US) representations, depicted as small purple circles, corresponding to rewarding stimuli such as juice or food pellets. The model assumes that extinction engrams are formed within this population, utilizing the reward circuitry to suppress fear responses and highlighting the overlap between extinction learning and reward processing.

- **Negative BLA Population**: This population consists of neurons that respond to negative or threat-related stimuli and encode fear associations. In Figure 1b, the negative BLA population is indicated by the red circle. Existing US representations within this population, also depicted as small purple circles, correspond to aversive stimuli such as foot shocks or puffs of air. The model posits that fear acquisition strengthens the associations between threat-related cues, contexts, and the US representations within the negative BLA population.

The cue input connects directly to both BLA populations via distinct positive and negative cue pathways from the sensory cortices. The context input reaches the BLA via distinct positive and negative context pathways from the hippocampal CA1 region, through the infralimbic (IL) and prelimbic (PL) regions of the ventromedial prefrontal cortex (VmPFC) respectively. The context pathways include:

- **Infralimbic (IL) Pathway (Positive Context Pathway)**: Connects the Hippocampus CA1 region to the positive BLA population via the IL, facilitating fear suppression.

- **Prelimbic (PL) Pathway (Negative Context Pathway)**: Connects the Hippocampus CA1 region to the negative BLA population via the PL, promoting fear expression.

The balance of activation between the positive and negative BLA populations in any given experimental trial, activated by connections from the input cue and context, determines the model's predicted fear response. ConFER computes this fear response by summing the scaled activations of both BLA populations in a given trial, and applying a non-linear activation function (tanh) to produce a final response within a bounded range [−1,+1]. Negative values indicate a fear response, while positive values indicate an appetitive response.

## O2. Weight Update Mechanism and Extinction Engram Formation

During each trial, ConFER selectively strengthens connections between the cue and context inputs and the active BLA engram, based on the presence or absence of a US. If a US is present, the corresponding US engram in the BLA is activated (Figure 1b, yellow-filled circle). Both cue–BLA and context–BLA connections with the active US engram are strengthened. The cue's connections with the active BLA engram are strengthened to a greater degree than the context's connections, reflecting the model's assumption that cue-based associations form faster than context-based associations.

If no US is present, ConFER checks whether the input cue and context evoke an emotional response, indicating an existing, learned association in the BLA. If such a response exists, ConFER deploys the extinction paradigm to neutralize this activation by activating and strengthening an extinction engram in the opposite-valence BLA population. An existing fear association due to negative BLA activation requires the formation of an extinction engram in the positive BLA, whereas an existing rewarding association is extinguished by forming an extinction engram in the negative BLA. Extinction engrams are specific to each cue and are formed by recruiting a random set of neurons from the appropriate BLA population. Once formed, the same set of extinction engram neurons is always activated to extinguish that specific cue. One of the core assumptions of ConFER is that it carries out extinction exclusively via the context pathway, strengthening the association between the extinction context and the cue's extinction engram in the BLA. Cue–BLA weights are not updated in the absence of a US.

Importantly, the cue-specific extinction engram associated with a conditioned stimulus (CS) can only be activated in the presence of that CS. An extinction context alone, despite any previously learned associations with an extinction engram, cannot activate it unless the associated CS is also present. During extinction learning, the CS activates the neurons in its cue-specific extinction engram, enabling synaptic plasticity between the extinction context and the extinction engram. Over repeated extinction trials, these synaptic connections gradually strengthen until the extinction context can sufficiently activate the positive BLA, suppressing the fear response.

Because extinction engrams can only be activated in the presence of their associated CS, the context–extinction engram associations learned during extinction are only utilized when that specific CS is present. This means that for each new cue extinguished in the same extinction context, the context must form a new association with that cue's own extinction engram to suppress fear. Previously learned associations between the extinction context and other extinction engrams cannot be leveraged unless their corresponding cues are present to activate them. As a result, fear extinction remains cue-specific in ConFER—each CS has its own extinction engram and must independently strengthen its context associations.

**O3. Temporal Dynamics of Contextual Associations**

A core assumption of ConFER is that all context–BLA associations naturally decay over time between trials, simulating forgetting processes, while cue–BLA associations are more stable. At the start of each trial, an exponential decay function is applied to all existing context–BLA connection weights, based on the time elapsed since the previous trial. This decay mechanism ensures that recent contextual associations retain more influence on fear responses, while older associations weaken progressively. The decay function is applied uniformly across both positive and negative BLA pathways and reflects the idea that emotional associations with a context weaken continuously over time, even when no CS is presented.

**O4. Capturing Return of Fear Phenomena Post Fear Extinction**

*Spontaneous Recovery*

Spontaneous recovery refers to the return of fear associated with the fear-extinguished cue within the extinction context itself following a substantial delay[18]. The ConFER model explains this phenomenon through the decay of context associations relative to the stability of cue associations. Over time, the association between the extinction context and the extinction engram weakens, allowing the cue-driven negative BLA activation to dominate, leading to a gradual re-emergence of the fear response.

*Fear Renewal*

Fear renewal occurs when a previously extinguished cue elicits a fear response in any context other than the one in which extinction took place[16]. In ConFER, extinction learning relies on context-specific associations with the extinction engram in the positive BLA. When the cue is presented in a different context that lacks these associations, the extinction engram is not activated. Without activation of the extinction engram, the negative BLA population becomes dominant, leading to the return of fear.

**O5. Summary of Model Assumptions**

By modeling the context-dependent return of fear after extinction, ConFER provides a mechanistic framework that highlights the role of contexts in guiding future experimental research and improving therapeutic interventions for anxiety disorders. The model fundamentally distinguishes between cue-specific and context-specific pathways, incorporating differences in their learning rates and susceptibility to decay, and emphasizes the formation of cue-specific fear extinction engrams within the positive BLA population. These core assumptions allow ConFER to capture key context-dependent return-of-fear phenomena and generate a novel prediction that counterconditioning may be more effective than extinction at preventing the return of fear.

The model is not intended to replicate the brain's fear circuit exactly, as the precise components and interactions remain an area of ongoing research with no established consensus. Instead, it offers a simplified, biologically plausible framework to explore mechanisms underlying context-dependent fear relapse.

**Results:**

The following experimental simulations sampled cues (CSs) and contexts from predefined sets of five 3-dimensional cue vectors and five 3-dimensional context vectors, respectively. Fear-inducing stimuli – representing painful shocks, puffs of air to the face, foul odor, etc. – and positive-reward stimuli (USs) – juice, sugar, etc.– were selected from predefined sets of four 4-dimensional negative and positive US vectors. The memory engrams corresponding to these USs are represented by the purple circles encompassing 4 neurons each in the negative and positive BLA respectively in Figure 1. Each negative US activated a predefined set of 4 neurons within a pool of 16 negative BLA neurons, while each positive US activated a predefined set of 4 neurons within a pool of 16 positive BLA neurons. The filled yellow circle in the negative BLA depicts an active fear-inducing stimulus in Figure 1. The indexing in the results corresponds to the index of the selected CS, US, and context. In the results, CSs and USs are numbered from 1 to 5 respectively, and the contexts are labeled from A to E. These cue and context inputs were chosen randomly to represent different phenomena, ensuring that the results were not influenced by specific input selection.

We assigned different learning and decay rates to the cue-BLA and context-BLA pathways to reflect the differing speeds at which emotional associations form and fade in cues and contexts during a cued fear conditioning paradigm. In ConFER, cue-BLA associations strengthen more quickly than context-BLA associations and do not decay over time. In contrast, context-BLA associations weaken exponentially as time passes. Further details are provided in the Methods.

**R1. Simulating Fear Acquisition and Extinction across Experimental Trials**

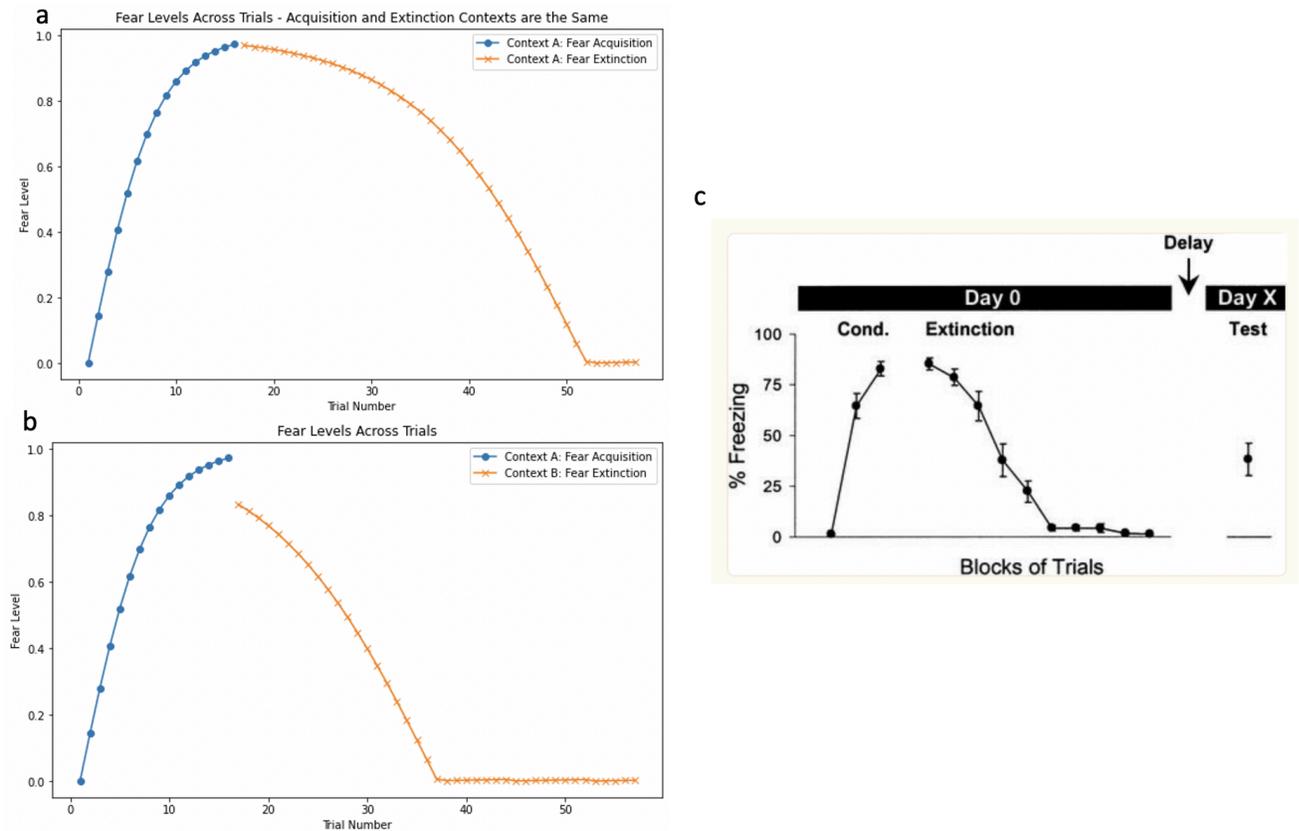

Figure 2: **Fear Acquisition and Extinction Learning Curves** (a and b) Model predictions for fear acquisition learning across trials (*in blue*) and extinction learning across trials (*in orange*). (a) Acquisition

and Extinction take place in the same context (b) Extinction takes place in a different context (c) Acquisition (Cond.) and Extinction curves from the Quirk (2002) fear conditioning study in rats.

*Fear Acquisition*

We paired a neutral cue (CS-1) with a shock (negative US-1) across 16 experimental trials in the acquisition context (Context A). ConFER's predicted fear level for each trial is shown in the blue curves in Figures 2a and 2b. The model predicts that fear acquisition follows an increasing, concave-down trajectory, with fear levels rising rapidly in the early trials and slowing as the response approaches its maximum (Figure 2a, blue curve). This pattern arises from ConFER's use of the tanh function to compute fear responses, which inherently produces a concave-down relationship between net negative BLA activation and fear levels across trials. This approach reflects the animal's initial surprise at the sudden pairing of a neutral cue and context with a negative US, followed by slower learning as the animal begins to anticipate the association. These predictions align with findings from studies on fear acquisition in rats, such as the Quirk (2002) study on fear learning, extinction, and spontaneous recovery[16] (Figure 2c).

In this experiment, both CS-1 and Context A were assumed to have no prior emotional associations with the positive or negative BLA populations, mimicking an animal encountering both for the first time. Consequently, the initial connection strengths between CS-1 and negative US-1, as well as between Context A and negative US-1, were set to zero. With each shock trial, the connection weights between CS-1 and negative US-1 and between Context A and negative US-1 were incremented, with cue weights increasing more substantially than context weights.

Since neither CS-1 nor Context A had preexisting positive BLA connections, the positive BLA was not activated during fear acquisition. Consistent negative BLA activation led to a consistent fear response across trials. Computing the final fear response using a tanh function of the net BLA activation resulted in the concave-down learning curve.

*Fear Extinction*

We simulated fear extinction in two contexts: the acquisition context (Context A) and a novel context (Context B). While the extinction learning curves were similar in both contexts, ConFER predicted a higher level of initial fear in the acquisition context, requiring more extinction trials to fully extinguish the fear response compared to the novel context. Notably, across both contexts, extinguishing the fear response required more trials than acquiring it within ConFER.

*Fear Extinction in the Acquisition Context*

Following fear acquisition, we presented ConFER with the fear-conditioned stimulus (CS-1) in the original acquisition context (Context A), but without pairing it with a shock. We provided 41 extinction trials, but 36 trials were sufficient to fully extinguish the fear response in the model (Figure 2a). The resulting extinction curve followed a decreasing, concave-down trajectory, reflecting slower extinction learning in the initial trials, with more rapid learning as the fear response approached its minimum. The shape of the extinction curve stems from ConFER's use of the *tanh* function to predict fear responses, where decreasing negative BLA activation produces a decreasing, concave-down trajectory in fear levels. This pattern is consistent with extinction curves reported in empirical studies, such as Quirk (2002)[16] (Figure 2c), and suggests the animal's hesitation to lose an acquired fear until it gathers enough evidence that it is indeed safe to stop being scared.

When a cue is presented in the absence of a US, ConFER activates the corresponding negative engrams in the BLA based on existing associations with the cue-context pair. In this experiment, CS-1 and Context

A had pre-existing negative associations from prior pairing with negative US-1, and since the experiment was conducted immediately after acquisition, the context associations had not yet decayed.

With no US present, ConFER needed to activate an extinction engram to neutralize the CS-1–shock association. During the first extinction trial, ConFER created a new extinction engram by recruiting 4 random neurons from a pool of 16 positive BLA neurons. Across subsequent extinction trials, the connection strength between the context and the extinction engram increased, while the cue–extinction engram connection remained unchanged. This reflects ConFER's assumption that extinction operates through the context pathway, neutralizing the existing cue-shock association without unlearning it.

Extinction continued until the net positive BLA activation from the extinction engram matched the net negative BLA activation from the learned fear association via both the cue and context pathways. Once this balance was achieved by trial 36, the fear response remained at zero for the remaining trials. Notably, the extinction engram formed for CS-1 would be re-used if CS-1 required re-extinction in future experiments.

ConFER predicts that fear extinction requires significantly more trials than fear acquisition because extinction relies solely on the slower context pathway, while fear acquisition utilizes both the rapid cue pathway and the slower context pathway. This difference in learning rates explains why it takes more trials to generate sufficient positive BLA activation to neutralize existing negative BLA activation.

*Fear Extinction in a Novel Context*

The extinction process in the novel context (Context-B) followed the same mechanisms as in the acquisition context. However, Context-B was assumed to be neutral, with no pre-existing emotional associations. This resulted in a lower initial fear level at the start of extinction compared to Context A. In Context A, the fear response was driven by the combined negative BLA activations from both the CS1–negative US-1 and Context A–negative US-1 connections. In contrast, in Context-B, only the CS1-negative US-1 connection contributed to the fear response. Consequently, fewer trials were needed to fully extinguish the fear response. Although ConFER was provided with 26 extinction trials, it required only 21 trials to achieve complete fear extinction. The extinction learning curve associated with a novel context is shown in Figure 2b.

**R2. Simulating Fear Renewal Across Different Context Combinations**

We also used ConFER to simulate the renewal of fear across different context combinations. In these simulations, context combinations are described as a three-letter sequence representing the acquisition, extinction, and testing contexts, respectively. For example, ABC renewal refers to an experiment where fear is acquired in Context A, extinguished in Context B, and tested for renewal in Context C.

We tested ConFER's predictions across five context combinations. In the first set of simulations, fear acquisition and extinction occurred in two different contexts (Contexts A and B), and renewal was evaluated in three combinations: ABA, ABB, and ABC. In the second set, acquisition and extinction occurred in the same context (Context A), and renewal was tested in two combinations: AAA and AAB.

The results were compared to empirical findings from animal studies[16,19,20]. In all simulations, the level of fear renewed was evaluated on the first trial in the testing context, with negligible delay between acquisition, extinction, and testing stages.

*Fear Renewal when Fear Acquisition and Extinction Occur in Different Contexts*

*ABA Renewal*

In this experiment, fear was acquired in Context A and subsequently extinguished in a different context, Context B. To test for ABA renewal, ConFER was reintroduced to Context A and presented with a single trial of the conditioned stimulus (CS-1), unpaired with any unconditioned stimulus (US). The model predicts a complete return of fear, with the renewed fear level matching the maximum fear acquired during the original acquisition phase. This prediction is shown as a red bar in Figure 3a, alongside the corresponding empirical result depicted in Figure 3b. ConFER successfully replicates the empirical finding that returning an animal to the acquisition context after extinction in a different context results in a full return of fear.

This complete return of fear is driven by ConFER's context-specific extinction mechanism. As depicted in Figure 3f, extinction operates via the context pathway, allowing Context B to form connections with and activate an extinction engram in the positive BLA population. However, upon returning to Context A, the extinction engram is no longer activated, as Context A has no associations with it (Figure 3g). Instead, Context A retains its associations with the negative US-1 engram formed during the acquisition phase.

Additionally, CS-1 maintains a strong connection with the negative US-1 engram in the negative BLA, as these cue-based connections remain stable over time. With both Context A and CS-1 activating the negative BLA and no activation of the positive BLA, the model's net BLA activation output, processed through the tanh function, results in a strong fear response.

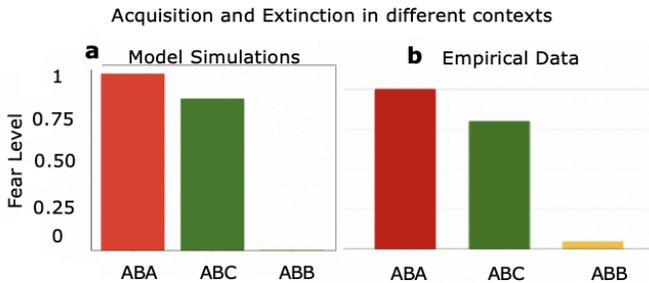
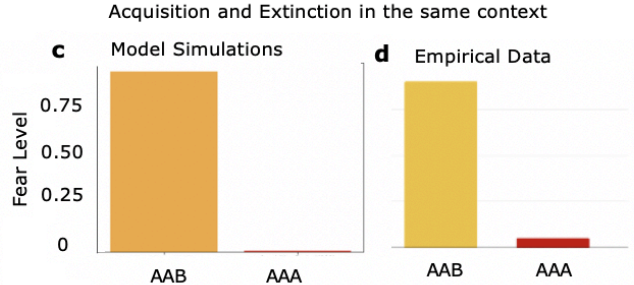
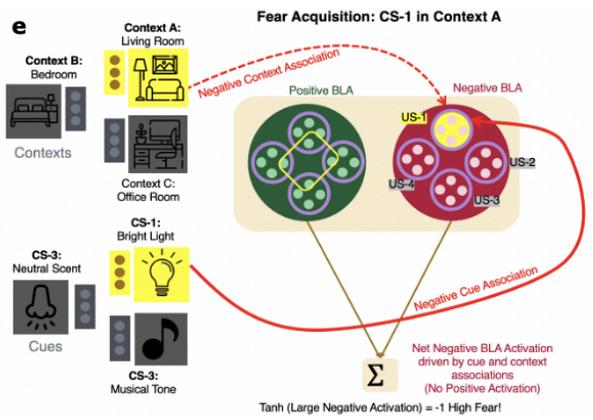
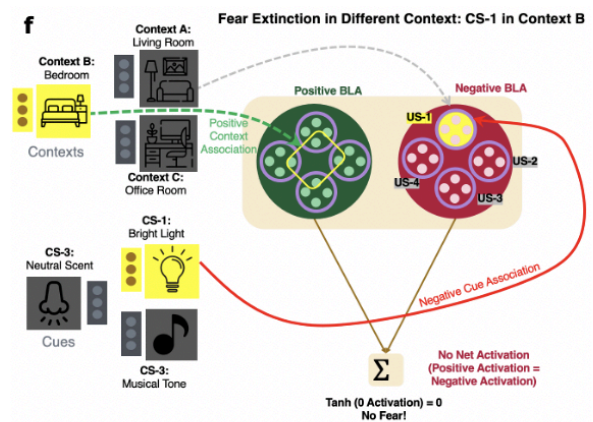
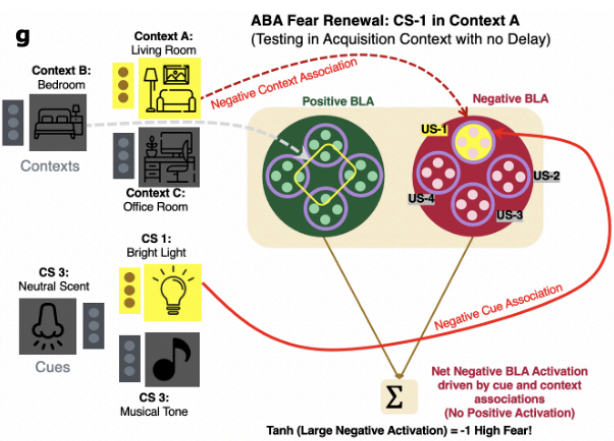
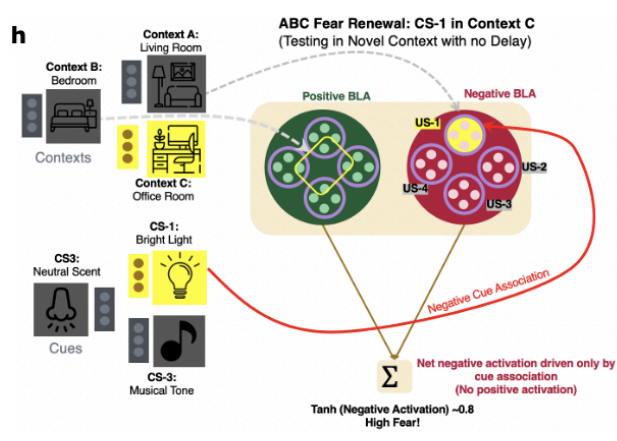
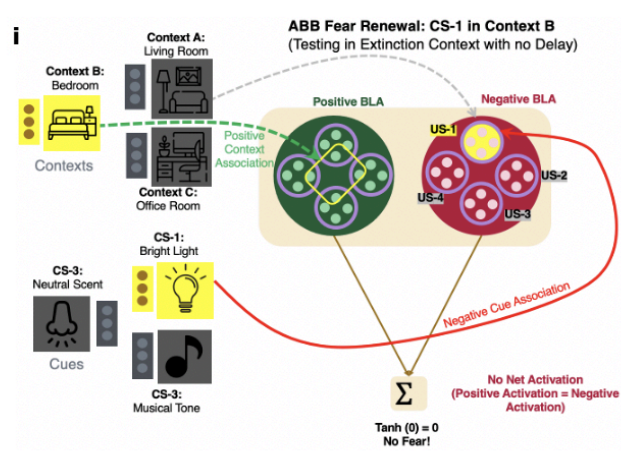

Figure 3: **Fear Renewal Across Different Context Combinations.** The charts on the left contain ConFER's predictions, and the charts on the right reflect corresponding empirical findings from Bouton (2004), Bouton and Bolles (1979a), and Bouton and Ricker (1993). (a and b) Level of fear renewed when acquisition and extinction take place in different contexts A and B, respectively. Bars show the level of fear renewed in the acquisition context (ABA Renewal in red), a novel context (ABC Renewal in green), and in the extinction context (ABB Renewal in yellow). (c and d) Level of fear renewed when acquisition and extinction take place in the same context A. Bars show the level of fear renewed in a novel context (AAB Renewal in yellow) and in the acquisition/extinction context (AAA Renewal in red). (e through f) Model mechanisms involved in fear acquisition in Context A (e), fear extinction in Context B (f), and fear renewal in Context A (*ABA Renewal*) (g), Context C (*ABC Renewal*) (h) and Context B (*ABB Renewal*) (i). Dotted associations are susceptible to temporal decay while solid associations are not. Thickness of arrows depicts strengths of connection weights. Red arrows show negative BLA associations, while green arrows show positive BLA associations. Greyed out arrows depict existing associations that are inactive for the presented cue-context pair. Fear acquisition involves strengthening cue (CS-1) and acquisition context (Context A, Living Room) associations with the negative US present (US-1) (e), whereas fear extinction only involves strengthening of the extinction context (Context B, Bedroom) association with the associated extinction engram (yellow rectangle) (f). ABA Renewal reactivates CS-1 and Acquisition Context associations with US-1 leading to a full return of fear (g). ABC Renewal works only through the CS-1 association, with neither the positive extinction context association nor the negative acquisition context association present, leading to a slightly decreased return of fear (h). ABB Renewal reactivates the extinction engram through the extinction context pathway which neutralizes the negative CS-1 association with US-1 (i).

*ABC Renewal*

In the ABC renewal experiment, CS-1 was presented in a novel context, Context C, after fear acquisition in Context A and extinction in Context B. ConFER predicts a substantial, though incomplete, renewal of fear in Context C, consistent with empirical findings and lower than the complete renewal predicted in the ABA condition. The model's prediction is shown as the green bar in Figure 3a, with the corresponding empirical result in Figure 3b.

The difference between ABA and ABC renewal stems from a key distinction: in ABC renewal, only CS-1 retains connections with the shock engram in the negative BLA (Figure 3h), while the context-based associations formed during acquisition (Context A) no longer contribute to negative BLA activation. This results in slightly lower negative BLA activation in the ABC condition. However, this reduction is modest, as cue-based associations remain stronger due to their faster learning rate compared to context associations.

*ABB Renewal*

In the ABB renewal experiment, CS-1 was presented in the extinction context (Context B) after fear acquisition in Context A and extinction in Context B. ConFER predicts almost no renewal of fear in this context, as shown by the yellow bar in Figure 3a, with the corresponding empirical result in Figure 3b. The model's prediction aligns with empirical findings, demonstrating that fear does not significantly return in the extinction context immediately after extinction. This outcome occurs because Context B maintains strong connections with the extinction engram in the positive BLA, which neutralizes the association between CS-1 and the shock engram in the negative BLA (Figure 3i). This positive context association is prone to decay over time, but since testing occurs immediately after extinction, these associations remain intact, preventing the return of fear.

*Fear Renewal when Fear Acquisition and Extinction Occur in the Same Context*

*AAB Renewal*

In the AAB renewal experiment, fear was acquired and extinguished in the same context (Context A), and renewal was tested in a novel context (Context B). ConFER predicts a significant return of fear in Context B, as shown by the yellow bar in Figure 3c, with the corresponding empirical result in Figure 3d. This renewal occurs because the extinction engram formed during extinction has connections only with Context A and no associations with Context B. The model captures the similarity between AAB and ABC renewal, where fear returns in a novel context, but not to the same extent as in ABA renewal, where the acquisition context itself contributes to the fear response.

*AAA Renewal*

In the AAA renewal experiment, fear acquisition, extinction, and renewal testing all occurred in the same context (Context A). ConFER predicts no fear renewal in this condition, as shown by the red bar in Figure 3c, with the corresponding empirical result in Figure 3d.

The mechanism is identical to ABB renewal: since the testing context and extinction context are the same, the extinction context's strong association with the positive BLA persists during testing, preventing fear renewal. The key difference, which is not immediately apparent, is that because extinction occurred in Context A, both cue-shock and context-shock associations contributed to the negative BLA activation that was neutralized during extinction. This led to a stronger extinction engram activation in the AAA condition compared to ABB, making the extinction engram more resistant to decay over time.

**R3. Spontaneous Recovery**

Spontaneous recovery refers to the return of fear after extinction, occurring in the extinction context following a prolonged delay. In this experiment, fear acquisition took place for CS-1 in Context A, followed by extinction in the same context. Testing was conducted in Context A after varying delay periods post-extinction, ranging from immediate testing to 21 days. Unlike AAA renewal, where testing occurs immediately after extinction, spontaneous recovery introduces a delay period, allowing context–extinction associations to weaken over time.

ConFER predicts that the fear gradually returns over time, with a sharp increase in the first few days, stabilizing as the fear response reaches the level following the initial acquisition phase. This concave-down, increasing trajectory of fear recovery is shown in Figure 4a. ConFER's predicted trajectory closely matches the empirical findings from the Quirk (2002) study of spontaneous recovery[18], depicted in Figure 4b. In ConFER, the complete return of fear is simulated over a period of 21 days by adjusting the decay rate of context-emotion associations, allowing them to diminish entirely over approximately 500 time units, corresponding to 500 hours or 21 days.

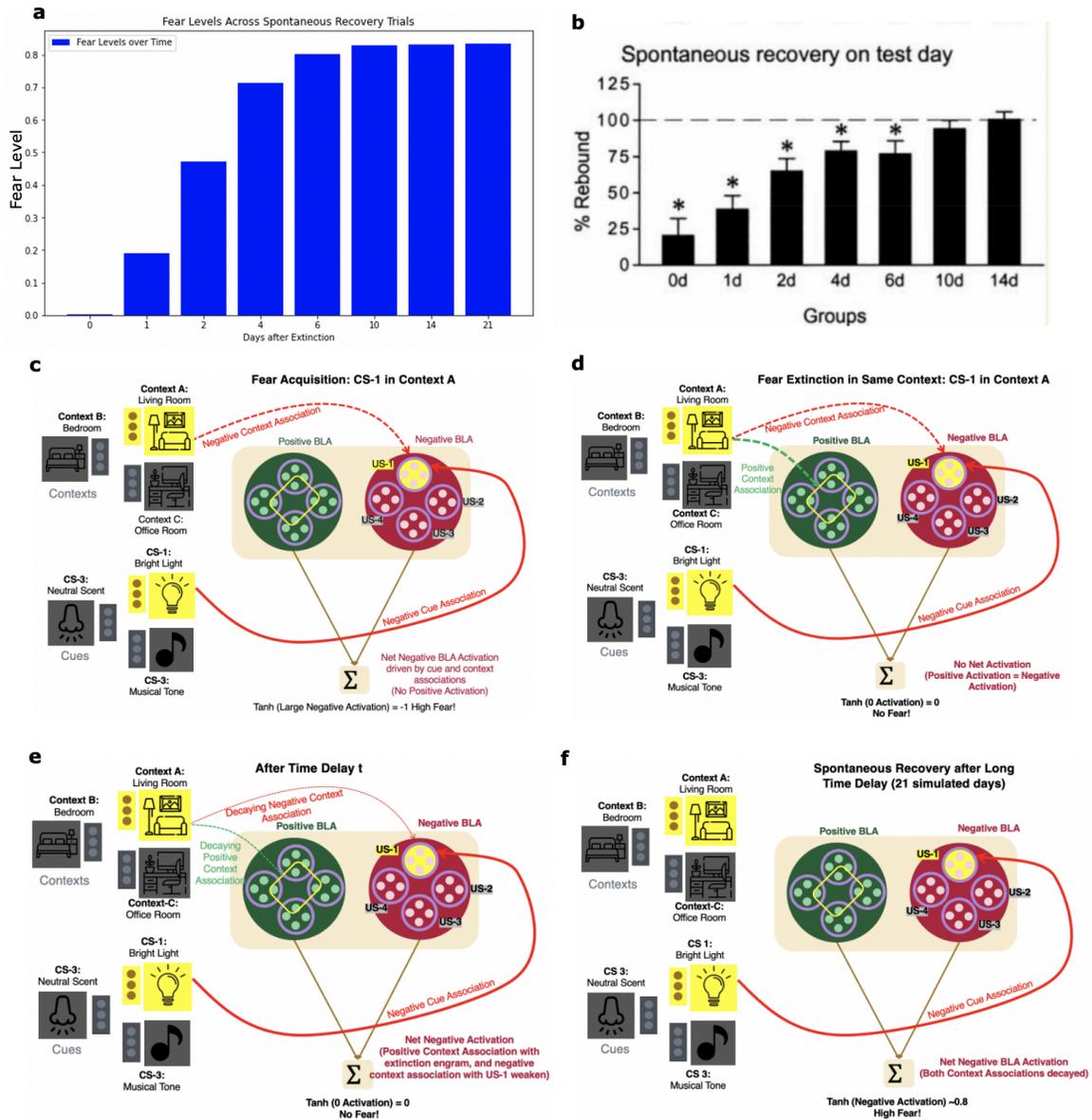

Figure 4: **Spontaneous Recovery.** (a) ConFER's predictions for the amount of fear that returns after prolonged delays after extinction, with the delay measured across 21 simulated days. (b) Empirical findings on the return of fear following fear extinction in rats across 14 days from Quirk (2002) show a similar learning curve over time. (c through f) Model mechanisms involved in fear acquisition in Context A (c), fear extinction in Context A itself (d), and exponentially increasing spontaneous recovery in Context A following a time delay t (e) and complete spontaneous recovery in Context A after a 21 day simulated delay within ConFER (f). Dotted associations are susceptible to temporal decay while solid associations are permanent. Thickness of arrows depicts strengths of connection weights. Red arrows show negative BLA associations, while green arrows show positive BLA associations. Fear extinction in the acquisition context involves the acquisition context forming new positive associations with the extinction engram to neutralize existing cue and context associations with US-1 (d). With a time delay t, context associations decay as indicated by the thinning dotted lines, while the negative cue association begins to dominate (e).

Day 0 of testing spontaneous recovery, immediately after extinction, corresponds to the AAA renewal condition, where the Context A—extinction engram association remains intact and prevents fear renewal. However, as time passes, these context–extinction associations begin to decay, following a concave-down exponential pattern, with rapid decay initially and slower decay over time (Figure 4e). As the connection between Context A and the extinction engram weakens, positive BLA activation diminishes, while the stable CS-1—shock association continues to activate the negative BLA. This imbalance leads to a progressive return of fear in the extinction context over time, until a near-complete return of fear occurs after a prolonged delay (Figure 4f).

Although the Context A—shock engram association formed during acquisition also decays over time, its impact on spontaneous recovery is minimal. The cue-based CS-1—shock association remains stable and contributes the majority of negative BLA activation, driving the return of fear in the extinction context. The use of the tanh function ensures that the fear response remains bounded, preventing it from exceeding the maximum fear level observed during the original acquisition phase.

### R4. Comparison of Fear Renewal and Spontaneous Recovery Following Counterconditioning vs. Extinction

We conducted a series of counterconditioning experiments to assess the long-term efficacy of counterconditioning compared to extinction in preventing the return of fear. These experiments examined fear renewal across the ABA, ABB, and ABC conditions. We did not include conditions where acquisition and extinction occurred in the same context, as ConFER predicts equivalent outcomes for ABB and AAA (test context = extinction context), and for ABC and AAB (test context ≠ extinction context). Additionally, we evaluated spontaneous recovery in the extinction context after a 21-day delay.

The counterconditioning experiments began with a fear acquisition phase, consisting of 16 presentations of CS-1 paired with a shock (negative US-1) in Context A. Counterconditioning was then performed in a different context (Context B), where CS-1 was paired with a positive reward (positive US-2). ConFER required 9 trials of CS-1 paired with positive US-2 to sufficiently neutralize the fear response without reversing the reward contingency (Figure 5a).

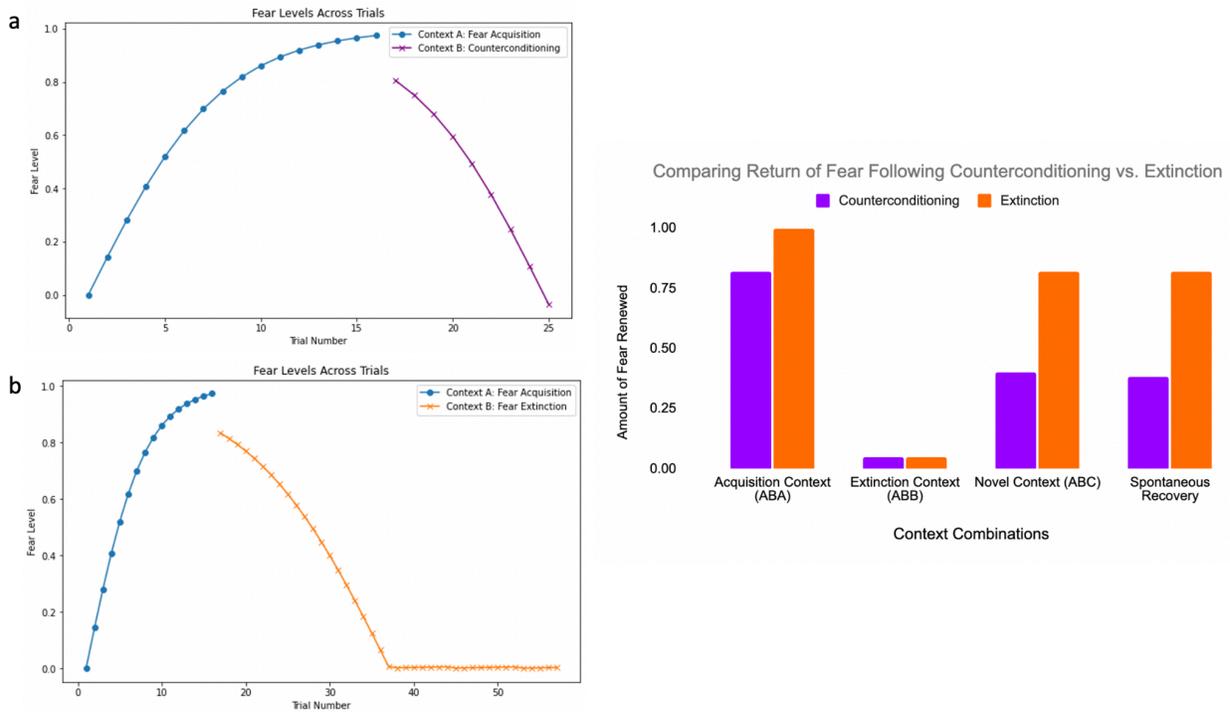

Figure 5: **Counterconditioning**. (a) Model predictions of levels of fear during Fear acquisition (in blue) and counterconditioning (in purple) across experimental trials. (b) Model predictions of levels of fear during Fear acquisition (in blue) and extinction (in orange) across experimental trials. (c) A comparison of the amount of fear that returns following counterconditioning (purple bars) and extinction (orange bars). The first 3 sets of bars represent the immediate return of fear via fear renewal in the acquisition, extinction, and novel contexts, respectively. The final pair of bars represents the return of fear in the extinction context after a 21-day delay via spontaneous recovery.

Counterconditioning operates through a separate mechanism compared to extinction in ConFER. While extinction relies solely on the context pathway to neutralize fear, counterconditioning allows the fear-conditioned CS-1 to form new associations with positive US-2. Moreover, Context B also develops associations with positive US-2. Together, CS-1 and Context B engage the same learning mechanisms used during fear acquisition during counterconditioning but with opposite valence. As a result, the faster learning cue pathway contributes to the reduction of fear, requiring fewer trials to achieve fear neutralization than in fear extinction (Figure 5b).

Following counterconditioning, we assessed the return of fear in Context A (ABA), Context B (ABB), and a novel context, Context C (ABC). ConFER's predictions for fear renewal in each condition are shown as blue bars in Figure 5c, with the corresponding renewal following extinction shown in red.

ConFER predicts that counterconditioning leads to a slightly reduced fear renewal in the acquisition context (Context A) compared to extinction (ABA). When ConFER re-encountered CS-1 in Context A following counterconditioning, the associations formed between Context-B and the positive extinction engram were lost. However, the CS-1–positive US-2 association formed during counterconditioning remained intact. Since ConFER was trained only to neutralize the fear response without reversing the reward association, the positive BLA activation associated with CS-1 was modest, allowing a large return of fear, albeit less than post-extinction.

In the counterconditioning context (Context B), there were no significant differences in the levels of fear renewal predicted between counterconditioning and extinction. Extinction neutralized fear through context associations with the positive extinction engram, while counterconditioning involved associations between both the cue and the context with positive US-2. Although the underlying mechanisms differ, both strategies prevented fear renewal when tested in the extinction context immediately after training, as the extinction associations remained intact.

ConFER also predicts that fear renewal in the novel context (Context C) is significantly lower following counterconditioning compared to extinction (ABC). After counterconditioning, although the associations with Context B and positive US-2 were inactive in the novel context, the CS-1–positive US-2 association provided sufficient activation of the positive BLA to compete with the existing CS-1–negative US-1 association from the acquisition phase. Nonetheless, as neutralization involved both CS-1 and Context B associations with positive US-2, a partial return of fear still occurred.

Lastly, we assessed spontaneous recovery in the counterconditioning context (Context B) after a 21-day delay. ConFER predicts that the return of fear following counterconditioning is less than half of the near-complete return it predicts after extinction (Figure 5c). This difference arises because, although the Context B–positive US associations formed during counterconditioning decayed over time, the Cue-2–positive US association remained intact, as cue-based associations are more resilient to decay. In contrast, extinction, which relied exclusively on the context pathway, led to the complete decay of all positive BLA associations over the 21-day period. As a result, the fear response after extinction fully re-emerged, whereas counterconditioning preserved a degree of fear suppression due to the enduring cue-positive US association.

**Discussion**

This section provides a comprehensive analysis of ConFER's contributions to understanding fear extinction and relapse, while situating the model within broader theoretical and clinical contexts. We begin by examining how ConFER integrates with existing extinction literature and highlights its unique mechanistic contributions to context-dependent fear relapse (D1). We then explores how ConFER bridges frameworks of associative learning, drawing distinctions between simple and configural approaches (D2). Next, we delve into the implications of ConFER's predictions for counterconditioning, proposing testable hypotheses with potential for improving clinical interventions (D3). We also address improving the clinical relevance of ConFER by considering extensions to simulate contextual conditioning and remote memory consolidation in fear conditioning (D4). Finally, we discuss the limitations of the current model and outline future directions(D5).

**D1. Situating Model Architecture and Mechanisms in Extinction Literature**

In this study, we developed a neurally constrained computational model, ConFER, that builds on the extensive literature on extinction from behavioral and neuroscientific perspectives[21] and contributes a novel mechanistic understanding of context-dependent relapse of fear. Our unique contribution is a simple computational model that preserves the fundamental structure of the tripartite fear circuit—comprising the hippocampus, amygdala, and medial prefrontal cortex—while mechanistically linking this circuit to recent findings that fear extinction engrams form by recruiting a random subset of neurons from the reward-responsive population of the basolateral amygdala (BLA), with these extinction engram cells being functionally interchangeable with the reward-responsive neurons[12], by incorporating existing knowledge that extinction learning and recall are highly context-dependent. ConFER replicates key experimental findings, including context-dependent fear renewal and spontaneous recovery. Below, we discuss how

ConFER integrates existing extinction theories, addresses fear relapse mechanisms, and informs clinical interventions for anxiety disorders.

One of the distinctive features of ConFER is that it has distinct pathways for cue and context processing pathways in a cued fear conditioning paradigm. We propose that the cue pathway forms fear associations more rapidly than the context pathway. During fear conditioning, the cue serves as the more salient predictor of the unconditioned stimulus, with sensory input from the cue directly projecting to the BLA, enabling rapid encoding of fear responses. In contrast, contextual information is processed through a slower route: sensory inputs from the environment project first to the CA1 region of the hippocampus, where the context is encoded, before projecting to the BLA via the Ventral Medial Prefrontal Cortex (VmPFC) to form a contextual fear association[22–24]. Specifically, the pathway from the downstream projections of the ventral hippocampus to the infralimbic cortex (IL), followed by projections of the IL into the BLA are implicated in fear extinction, while downstream projections from the ventral hippocampus to the prelimbic cortex (PL), followed by projections of the PL into the BLA are implicated in fear activation.

ConFER accounts for these distinct context-dependent fear and extinction pathways as depicted by the green and red context–BLA pathways respectively in Figure 1[24,25]. In addition to a difference in learning rates, ConFER also suggests that cue-BLA associations are more resilient to temporal decay, whereas context-BLA associations weaken over time. This reflects the idea that retaining fear responses to salient cues is critical for survival, while contextual information becomes less relevant with time. The persistence of cue-fear associations is evident in fear renewal phenomena, where fear responses return across different contexts, except the extinction context[16,19]. By contrast, spontaneous recovery offers insights into the temporal decay of context-extinction engram associations, as it involves no active experimental manipulation except for a time delay[18]. The full return of fear in the extinction context underscores that while the CS–fear engram association remains stable, the context–extinction engram association decays over time. By incorporating distinct pathways for cue and context associative learning with differential rates of learning and decay, ConFER is able to simulate context-dependent return of fear phenomena following extinction, expanding on the effects explained by existing models of associative learning by offering new insights into the nuanced roles of cue and context in fear renewal and spontaneous recovery.

ConFER also incorporates a memory engram-based approach to fear learning, extinction, and recovery. It posits that fear engrams are persistently stored in the negative BLA, consistent with findings that fear memory engrams persist throughout the lifespan and are essential for fear memory retrieval, as deactivation of the BLA results in the loss of these memories[26,27]. However, while the fear engram is persistent, it remains dynamic, as its activity can be modulated or silenced by surrounding neural circuitry[28,29]. In ConFER, fear extinction is represented by the formation and activation of an extinction engram in the positive BLA, which neutralizes the negative BLA activity during extinction. This is inspired by recent evidence showing that a random set of positive BLA neurons is recruited to form an extinction engram during extinction learning[12]. ConFER simulates fear extinction as the extinction engram in the positive BLA neutralizing the fear response in the negative BLA instead of an unlearning of the acquired fear that models of associative learning often suggest[8,9].

The mechanisms in our model align with the literature in several ways. Primarily, we model extinction as an inhibition of an existing learned association rather than as unlearning it altogether. The connectivity between regions in ConFER reflects empirical findings regarding the constituents of the fear circuit during extinction. For example, the BLA is necessary for fear extinction learning, with lesions to the BLA—but not to the central amygdala (CeA)—impairing the ability to learn extinction[30–34]. In ConFER, because the extinction engram forms within the BLA, damage to the BLA would disrupt the recruitment of extinction engram cells. However, damage to the CeA, which corresponds to the final fear output computation modeled via a tanh function, would only affect the generation of the fear response, not the extinction learning itself.

Another example concerns the role of the infralimbic cortex (IL) in the consolidation of extinction. IL lesions lead to increased freezing following extinction without affecting freezing during or after fear conditioning[35,36]. In ConFER, the IL is incorporated within the positive context pathway from the hippocampus CA1 to the extinction engram in the BLA. Disruptions to this IL pathway in our model would prevent the strengthening of context-extinction associations, which activate the extinction engram, thereby impairing extinction consolidation. However, because fear conditioning utilizes a separate mechanism involving the cue pathways and the context-negative pathway from the prelimbic cortex (PL), disrupting the IL pathway would not affect conditioned fear responses.

Additionally, research has shown that the BLA is essential not only for fear extinction but also for extinguishing rewarding associations[37]. This suggests that parallel mechanisms exist within the BLA to extinguish emotional responses of different valences. ConFER's mechanisms account for this by allowing both fearful and rewarding associations to be extinguished through the formation of extinction engrams in the BLA population of opposite valence to the original association.

However, it must be noted that ConFER is based on a highly simplified version of the fear circuit, retaining only key regions and connections. For example, in the fear circuit, there are bidirectional connections between the BLA and hippocampus, and between the BLA and the IL and PL. ConFER, however, only includes feedforward associations from the hippocampus to the BLA, and from the IL and PL to the BLA. The model also excludes downstream afferents of the CeA that facilitate the final fear response, such as the ventromedial periaqueductal gray (PAG), as well as the nucleus reuniens, which has bidirectional connections between the IL, PL, and hippocampus[21]. Building a more complex architecture with these additional regions could allow ConFER to capture more nuanced, experiment-specific extinction effects. However, doing so would go beyond the scope of this paper. Here, we aim to propose plausible mechanisms linking the tripartite fear circuit and the context dependency of extinction to recent findings regarding the extinction engram in the BLA's reward-responsive population. We hope that doing so could suggest new research avenues for improving the long-term efficacy of retroactive interference learning methods, such as extinction and counterconditioning, in preventing fear relapse. We have maintained simplicity in the model to introduce novel, easily interpretable mechanisms of context-dependent relapse.

### D2. Integrating ConFER into Existing Associative Learning Frameworks

Animals acquire a fear association to a cue independently of the context during fear conditioning. However, once the cue is extinguished in a specific context, the context becomes crucial in determining whether the animal exhibits a fear response or an extinction response to the now ambiguous cue. This suggests that during fear acquisition, fear is formed as simple associations between the individual elements of the cue and the features of the context with the US. In contrast, fear extinction relies on learning that integrates the cue-context configuration, suggesting that the configuration itself forms the associative link with the US. This distinction indicates that ConFER engages different associative learning frameworks—the simple associative theory, where individual elements form associations with the US, and the configural associative theory, where configurations of elements form associations— at different stages of fear learning and extinction recall[38].

Deterministic models of associative learning, such as the Rescorla-Wagner model and the Temporal Difference model of classical conditioning, align with the simple associative framework. In these models, the cue and context elements are treated collectively as a compound cue, with the most predictive elements increasing in associative strength relative to their association with the US[8,9]. In contrast, probabilistic models like the Latent Cause Theory add an additional layer of inference by using both cue and context features, along with the presence or absence of a US, to attribute experimental trials to different latent causes in the environment. In this framework, the combination of individual features infers underlying categories or causes[6,7]. ConFER integrates both simple and configural associative frameworks. When the US is present, ConFER operates within the simple associative framework, where both cue and context dimensions increase their associative strength through distinct pathways with different learning and decay rates. During extinction, however, ConFER shifts towards the configural associative framework. In this stage, it computes the net activation of the cue-context configuration and seeks to neutralize this activation during extinction learning. Extinction involves forming associations between the extinction context and a cue-specific extinction engram, which is more specific to the cue-context configuration than to either element independently. As a result, ConFER aligns more with the configural associative theory during extinction learning and recall.

A key feature of ConFER is its departure from a strict prediction error (PE) framework, distinguishing it from classic associative learning models such as Rescorla-Wagner and Temporal Difference (TD) models, where PE directly drives changes in associative strength. In these models, learning is traditionally governed by the degree of expectation violation: large associative updates occur with surprising outcomes, while expected outcomes yield smaller updates. In contrast, ConFER's learning rates are independent of prediction error and depend solely on the nature of the connection pathway—specifically, whether it is cue- or context-based. Extinction studies, including those by Quirk (2002)[18], suggest that a PE-based learning framework may be insufficient to fully capture extinction dynamics in behavior without an additional performance layer. For instance, during early extinction trials, the unexpected absence of the US generates high PE, yet animals show minimal reductions in fear responses, reflecting a slow behavioral shift. ConFER interprets this effect not as a slow learning process, but as a cautious performance strategy, wherein performance is represented by the tanh output of the net activation in the BLA for each trial. The use of a tanh function allows ConFER to simulate a performance-based response that prioritizes safety, yielding the gradual behavioral adaptation predicted during extinction. This adaptation arises from a concave-down decrease in fear responses as negative BLA activation decreases, effectively modeling the cautious modulation of fear during extinction.

### D3. Counterconditioning: Implications and Predictions

ConFER can also be used to generate new behavioral hypotheses based on its predictions in experimental manipulations. These hypotheses could inform the design of experimental paradigms aimed at improving our understanding of extinction recall in humans. One such hypothesis we explored in detail is that counterconditioning should be more effective than fear extinction at preventing fear relapse. ConFER predicts that while counterconditioning will not significantly reduce fear relapse in the original fear acquisition context, it will be far more effective at preventing relapse in a novel context. This is a key prediction, as overgeneralization of fear to stimuli and contexts not directly linked to the initial fear acquisition is a hallmark symptom of Post-Traumatic Stress Disorder (PTSD) and other anxiety disorders[39,40]. This difference between counterconditioning and extinction stems from the involvement of different pathways within ConFER to carry out each process. Fear extinction relies on the extinction context, with the extinction context forming and strengthening connections with the extinction engram. Not only is this connection prone to decay (resulting in Spontaneous Recovery), but it is specific to the context in which it is formed since the extinction engram can only be activated via a specific context association (leading to fear renewal). Counterconditioning on the other hand, primarily leverages the more permanent cue pathway, by forming a connection between the conditioned stimulus and a positive US. This association is less prone to decay (preventing complete spontaneous recovery), and is not context-specific (preventing renewal), since the positive response is activated by the cue. ConFER further suggests that counterconditioning should require fewer trials to neutralize the fear response compared to extinction, which is an essential distinction if this can minimize patient distress during exposure therapy. This is because of the reliance of counterconditioning primarily on the faster cue pathway as opposed to the context pathway, leading to faster learning of this new association.

Seminal work exploring mechanisms underlying renewal and spontaneous recovery following counterconditioning suggests that, like fear extinction, counterconditioning involves learning a novel stimulus association that suppresses the existing fear association without unlearning it, which aligns with ConFER's assumptions. Bouton and Peck[41] conducted two counterconditioning experiments to evaluate spontaneous recovery following counterconditioning. The first involved an aversive-to-appetitive transfer, in which a CS was first paired with a negative stimulus in phase 1, followed by a positive stimulus in phase 2. The second involved appetitive-to-aversive transfer, with a positive US pairing in phase 1 followed by a negative US pairing in phase 2. Across these experiments, they arrived at two major findings. First, they found that spontaneous recovery following counterconditioning resulted in the return of the learned phase 1 response, suggesting that the phase 2 learned association served to suppress the existing phase 1 association rather than erase it. They postulated that suppression was contingent on the temporal context in which counterconditioning was learned and consequently not retrieved when the temporal context changed. ConFER's predictions align with these findings since counterconditioning in the model yields a spontaneous recovery of the original learned association. When the counterconditioning context is different from the original acquisition context, both the CS and counterconditioning context form novel associations with the counterconditioning US. Although the new CS–counterconditioning US association is stable, the counterconditioning context–counterconditioning US association decays over time, which can be paralleled to a changing temporal context. With the decay of this association leading to a weaker activation of the engram associated with the counterconditioning US in the basolateral amygdala (BLA), the original cue association begins to dominate, albeit not completely.

Second, they found that this return of the phase 1 learned association was independent of the valence of this association, with the phase 1 response being recovered in both the aversive-to-appetitive and appetitive-to-aversive experiments. Since ConFER employs symmetric mechanisms across positive or negative learned associations with the BLA, it can reproduce this finding, with the original BLA activation dominating via the original cue–BLA association, irrespective of its valence.

In a different set of experiments, Peck and Bouton[42] also found that an immediate return to the phase 1 acquisition context following counterconditioning in phase 2 increased the phase 1 learned response and decreased the phase 2 learned response. ConFER predicts this pattern, with a return to the acquisition context reactivating the context associations with the original US while simultaneously resulting in a loss of counterconditioning context associations with the counterconditioning US of the opposite valence. On a related note, they found that a context switch following an initial aversive phase, where the CS is paired with a shock, reduces the retarded appetitive conditioning seen in the aversive context. ConFER also predicts this, as moving to a new context for counterconditioning results in the loss of the aversive context–aversive US connection formed during the aversive phase, leaving the counterconditioning phase with a lower BLA activation to neutralize.

Previous empirical exploration of counterconditioning as a potential alternative to yield better long-term outcomes in preventing the relapse of fear than extinction has produced mixed results, with relatively few studies explicitly making this comparison. Correia et al. built on the idea that extinction relies on reward circuitry to identify a BLA–Nucleus Accumbens (NAc) pathway between the amygdala and ventral striatum that is recruited during counterconditioning[43]. They found that counterconditioning yields superior suppression of the original fear association compared to extinction. Interestingly, they found that counterconditioning by pairing the fear-conditioned CS with a positive US yields the same outcome as counterconditioning by optogenetically activating the BLA–NAc pathway, suggesting a potential neurobiological mechanism for counterconditioning. Although the ConFER architecture does not include the NAc, both fear extinction and counterconditioning also rely on the reward circuitry within ConFER by activating the positive US engram in the positive BLA.

While counterconditioning in our experiment relies on presenting the CS paired with an opposite valence US, others leveraging modified versions of counterconditioning have also found counterconditioning to be superior to fear extinction in preventing relapse. For instance, Thomas et al. found that employing an action (instrumental lever pressing) to receive the positive US (chocolate milk) as opposed to freely providing the US during counterconditioning led to a more robust prevention of fear relapse[44]. Following contextual fear conditioning in rats, Anderson et al. added novel objects into the feared context during extinction, as opposed to the context alone, contingent on the assumption that novelty engages appetitive behavior[45]. They found that counterconditioning deployed by introducing these novel objects led to better suppression of fear than extinction. Future versions of ConFER incorporating instrumental conditioning and contextual conditioning could build on these findings to better understand the distinction between fear extinction and counterconditioning.

There have also been a few studies finding that counterconditioning leads to a stronger renewal of fear than extinction or is similar to extinction at best. Holmes et al. found that counterconditioning leads to a stronger ABA and ABC renewal than fear extinction[46]. They suggested that pairing a positive US with the fear-conditioned CS enhances the animal's ability to discriminate between the two distinct associations of the CS and reduce ambiguity. In this case, counterconditioning becomes entirely context-specific, and returning the animal to the original acquisition context (ABA) or a novel context (ABC) leads to a loss of the associations learned during counterconditioning. In ConFER, we propose that although extinction relies solely on such context-driven pathways, counterconditioning would also involve the more permanent cue pathway due to the preference for emotional vs. neutral outcomes in animals. In humans, Meulders et al. found that pairing a fear-conditioned joystick movement with no reward (extinction) as opposed to a monetary reward (counterconditioning) led to no differences in the relapse of fear[47]. In a study with college students extinguishing a fear association by pairing a CS with no reward as opposed to pairing it with a positive film clip, Van Dis et al. proposed a more nuanced argument[48]. They suggested that counterconditioning reduced the stimulus valence of how negative students found the CS much more than extinction, but did not significantly affect their fear responses during spontaneous recovery or renewal. In our model, this could point to the communication between the BLA and the central amygdala (CeA), in computing the final fear response from the BLA activation, with the BLA activation corresponding to the valence of the CS and the tanh function parallel to the CeA computing the actual fear responses. Van Dis et al. highlight considerations for how the threshold for generating a fear response from this activation might be set[48].

The proposed reliance of counterconditioning on the cue pathway in ConFER could point to effective avenues of preventing relapse in patients with hippocampal damage or impairments in context processing. Notably, PTSD has been linked to impaired contextual processing, suggesting dysfunction in this pathway[49,50], and individuals with PTSD have consistently shown smaller hippocampal volumes compared to controls[51]. Given that counterconditioning likely engages the faster, more resilient cue-based neural circuitry, it may offer a promising direction for improving clinical interventions. There remains a scarcity of studies directly comparing the long-term effects of counterconditioning and extinction in preventing fear relapse, and existing research has shown mixed results[46,47,52]. However, some initial efforts to incorporate counterconditioning in human exposure therapy have shown promising results[53], indicating that this approach warrants further exploration.

**D4. Additional Considerations for Improving ConFER's Clinical Relevance**

While ConFER presents interesting new mechanisms for context-dependent extinction of cued-fear conditioning, there are some aspects that require thoughtful consideration before a meaningful translation of these findings to exposure therapy is possible. An important consideration is the treatment of contexts within the model. ConFER currently attributes symbolic as opposed to distributed representations to contexts, where contexts exist as a predefined set of distinct, independent inputs available to the model as opposed to distributed representations with shared features, with the associations of one context generalizing to another based on overlap of features. With ConFER's emphasis on a cued fear conditioning paradigm, contexts carry out an accessory role in neutralizing or supplementing cue-driven emotional responses in the BLA. This is especially highlighted in ConFER's extinction mechanism, which involves the strengthening of the context association between the extinction context and the cue-specific extinction engram. However, stress and anxiety disorders can often stem from contextual conditioning, which involves associating a context or a situation, as opposed to a specific cue, with a US. While it is theoretically possible to model contextual conditioning within the model in its current form, fear extinction would require further consideration as extinction engrams are cue-specific in ConFER. Thoughtful incorporation of contextual fear conditioning within ConFER would also require additional considerations. One major issue is how the similarity of contexts encountered across different phases of contextual fear conditioning affects the fear responses to each context.

Building on the need to consider contextual conditioning, recent findings underscore the critical role of context encoding in shaping fear responses. Poor encoding of a context during conditioning—such as when there is insufficient time between placing the animal in a context and administering a shock (placement-shock interval, or PSI)—can result in a fear response that overgeneralizes to safe contexts and becomes resistant to extinction[54]. Post-conditioning exposure to the original context has been shown to mitigate this overgeneralization by allowing for better encoding of context-specific attributes, which helps attribute the fear memory to the correct context, making it easier to extinguish. Conversely, exposure to a similar but safe context post-conditioning can inadvertently shift the fear response to this safe context, complicating extinction further.

One model that accounts for many of these findings is the Bayesian model of contextual fear learning (BACON)[55], which explains how context encoding and similarity influence fear generalization. BACON simulates the formation of an initial representation of a context in the hippocampus, where the richness of the representation depends on the time available to sample the context's attributes. Following conditioning, the model evaluates new contexts by comparing their similarity to the original context representation. If the similarity is sufficiently high, the model recognizes it as the same context and enriches the existing representation by incorporating additional features. If the similarity is low, the model treats it as a distinct context and creates a new representation. Extensions of the BACON model, such as BACON-X[56] (which explores contextual fear extinction) and BACON-REM[57] (which addresses the formation of remote fear memories in the neocortex), further illustrate how the quality of context encoding impacts fear generalization. Poorly encoded contexts are more likely to share features with novel contexts, leading to overgeneralization of fear, whereas richly encoded contexts reduce generalization by making it harder to find close matches. The PSI directly influences the richness of these representations, highlighting the importance of sufficient time for encoding during conditioning. Future iterations of ConFER could incorporate these principles by encoding context information and computing similarities across encountered contexts to modulate fear responses based on the animal's certainty about being in a familiar context. By integrating these ideas, ConFER could extend its applicability to address the nuanced mechanisms of contextual fear conditioning, generalization and extinction.

Another critical consideration for increasing ConFER's clinical relevance is to incorporate the role of remote, systems-consolidated fear memories in anxiety and stress disorders, as opposed to recently acquired fears that the model currently evaluates. For example, PTSD is characterized by a persistence of symptoms for at least 1 month, with traumatic memories often persisting across a patient's lifetime[58]. Remote fear memories are stored and expressed differently from recent memories, with the longstanding hypothesis being that long-term consolidation of memories results in memories being shifted into the neocortex, strengthening intracortical connections while becoming independent of the hippocampus[59–61]. While these remote memories become more difficult to modify after consolidation, they become temporarily malleable in a short window following recall due to increased neuronal plasticity at this time via a process called reconsolidation updating[62]. Reconsolidation updating is garnering increased interest as a means of attenuating system-consolidated remote fear memories in anxiety and stress disorders[62]. Future versions of ConFER could incorporate remote consolidation and reconsolidation of fear memories to better bridge the gap between the novel mechanisms we propose and the application of these mechanisms to improve exposure therapy.

**D5. Limitations and Future Directions**:

A key limitation of ConFER is that it does not currently simulate a third type of relapse phenomenon known as reinstatement. Reinstatement occurs when fear returns in a safe context following a single presentation of the unconditioned stimulus (US). In ConFER, immediately after fear associated with a conditioned stimulus (CS) is extinguished, the extinction context retains a strong association with the cue's extinction engram in the positive BLA. A single presentation of the US in isolation would only instantiate a weak connection with the US engram in the negative BLA. Therefore, subsequent presentations of the extinguished CS in the extinction context would not lead to a return of fear, as the positive context association remains dominant.

Different approaches could be taken to extend the model to account for reinstatement. One approach could involve introducing a feedback connection from the BLA to the CA1 region of the hippocampus. This feedback would allow the context representation in CA1 to be modified based on the emotional activation of the BLA, enabling the context to switch from a safe to a dangerous representation upon activation of the negative US engram within that context. Another potential approach is to incorporate dynamic, flexible engram switching, whereby the context shifts its association from the extinction engram to the negative US engram when an emotional US is encountered. This assumes that emotional stimuli are more salient than safety associations, prompting the switch to prioritize vigilance in the presence of potential threats.

The current version of ConFER also does not take into account similarity between encountered contexts in determining the fear response within a context. ConFER, in line with animal findings, predicts that fear renewal occurs in any context other than the extinction context. Notably, the degree of fear renewal is influenced by the similarity between the test and extinction contexts—greater dissimilarity results in stronger renewal effects[63]. Since ConFER is a mechanistic model, looking to context discrimination mechanisms in the brain would be an important next step. This context-adaptive response likely engages pattern separation and pattern completion mechanisms to distinguish between extinction and test contexts. The dentate gyrus–hippocampal CA3 pathway is crucial for pattern separation, while the CA1 region and other medial temporal lobe structures are implicated in pattern completion[64–66]. However, ConFER currently treats all novel contexts as equally dissimilar from the extinction context, without calculating any context similarity. In future work, we aim to incorporate a similarity module to capture these more nuanced contextual fear responses.

Although ConFER currently employs multidimensional cues and contexts, it simplifies these representations by treating them as independent entities rather than adopting a distributed framework. As discussed in section D4, this implies that the fear learning associated with specific elements of one cue or context does not influence the associative strengths of similar elements in other cues or contexts. Presently, the multidimensionality serves primarily to demonstrate how distinct neurons from various US engrams are recruited at random to encode an extinction engram. In future iterations, we intend to adopt a distributed representation framework to enable generalization across shared features of cues and contexts, enhancing ConFER's flexibility and ecological validity.

Another direction to consider is the relationship between extinction mechanisms in Pavlovian and instrumental conditioning, where extinction research has identified several parallels[21]. While Pavlovian conditioning results in involuntary physiological responses to conditioned stimuli, instrumental conditioning involves reinforcing voluntary behaviors. Reinforced behaviors extinguished through instrumental conditioning have also been shown to demonstrate renewal and spontaneous recovery[67–69]. Despite these similarities, fundamental differences exist between the two types of conditioning. For instance, acquisition in instrumental conditioning is itself context-dependent, with reinforced behaviors not fully translating to novel contexts. Although extinction remains more context-dependent, as indicated by the renewal of behaviors in non-extinction contexts, the relapse effects in instrumental conditioning rely on different mechanisms and pathways.

Due to these differences, ConFER cannot be directly extended to explain relapse effects following extinction in instrumental conditioning. The extinction pathway in instrumental conditioning involves additional regions, such as the nucleus accumbens shell, which is activated via the IL[21]. Reinstatement in instrumental conditioning works through connections from the PL to the nucleus accumbens core[21]. While ConFER's mechanisms might provide initial insights into extinction in instrumental conditioning, explaining relapse effects in this context would require a more nuanced model incorporating additional regions and mechanisms.

Finally, although ConFER accounts for the time elapsed between trials, it does not yet incorporate within-trial temporal dynamics. The temporal difference (TD) model of classical conditioning has demonstrated that both the duration for which the CS is presented and the interval between CS and US presentations play key roles in shaping conditioning phenomena[8]. Incorporating such temporal variables within trials would allow ConFER to better simulate real-world conditioning processes.

**D6. Conclusion:**

We present a neurally constrained computational model that offers a novel framework for understanding the mechanisms underlying fear acquisition, extinction, and the context-dependent return of fear. By incorporating distinct cue and context pathways and simulating the formation of separate fear and extinction engrams in the basolateral amygdala (BLA), ConFER captures key experimental findings and generates testable hypotheses—particularly regarding the efficacy of counterconditioning over traditional extinction methods. These insights enhance our understanding of the neural underpinnings of fear learning and relapse, with significant potential for informing and improving clinical interventions for anxiety disorders.

**Methods:**

**M1. Model Architecture and Mechanisms:**

**Model Architecture**

The architecture of ConFER, illustrated in Figure 1b, is grounded in a simplified schematic of the fear circuit in the brain, shown in Figure 1a. ConFER takes as input a 3-dimensional cue vector and a 3-dimensional context vector, where each dimension represents distinct features of the cue and context, respectively. For example, a blinking yellow light as a cue might have dimensions representing its intensity, warmth, and blinking frequency, while the corresponding context dimensions might encode the room temperature, size, and furniture density. In this study, we utilized five distinct 3D cues and five distinct 3D contexts, but ConFER is designed to scale to accommodate larger sets as needed.

Within ConFER, the basolateral amygdala (BLA) module comprises two main populations of neurons:

- Positive-Stimulus-Responsive Population (green circle in Figure 1b): This population contains 16 neurons that respond to rewarding stimuli such as food, water, or social interaction.
- Negative-Stimulus-Responsive Population (red circle in Figure 1b): This population also contains 16 neurons that respond to aversive stimuli such as footshocks, foul odors, or puffs of air.

Each population is further subdivided into four memory engrams, each representing preprogrammed responses to specific unconditioned stimuli (US). These engrams correspond to distinct emotional representations. For example, within the negative population, one engram may encode a response to a footshock, while another encodes a response to a foul odor. These engrams are depicted as purple circles in Figure 1b, with the yellow-filled circle representing an active shock engram.

**Connection Matrices and Pathways**

ConFER uses connection matrices to store the strengths of associations between the input (cue and context), and the BLA neurons. These matrices correspond to the distinct cue and context pathways that connect sensory cortices and the hippocampus to the BLA, respectively.

The following matrices represent the connection weights that modulate the model's response to different cues and contexts:

- *Cue–Positive BLA Matrix:* This matrix corresponds to the strength of the *positive cue pathway* across different cues and reward representations, and stores connection strengths between cue features and neurons in the positive BLA population.
- *Cue–Negative BLA Matrix*: This matrix corresponds to the strength of *negative cue pathway* across different cues and threat representations, and stores connection strengths between cue features and neurons in the negative BLA population.
- *Context–Positive BLA Matrix:* This matrix corresponds to the strength of the IL pathway (positive context pathway), and stores connection strengths between context dimensions and neurons in the positive BLA population.
- *Context–Negative BLA Matrix:* This matrix corresponds to the strength of the PL pathway (negative context pathway), and stores connection strengths between context dimensions and neurons in the negative BLA population.

Each matrix has dimensions of 15×16, where the 15 input dimensions correspond to the features of five 3D cues or contexts, and the 16 columns represent the neurons in each BLA population. These matrices store the connection strengths between all input features across cues and contexts and all neurons in the BLA. These connection matrices are updated during each trial based on the inputs presented to the model (see Algorithm details in **Methods Section M3**).

**Temporal Dynamics of Contextual Associations**

ConFER assumes that all context–BLA associations naturally decay over time between trials, simulating forgetting processes. At the start of each trial, an exponential decay function is applied to all existing context–BLA connection weights, proportional to the time elapsed since the previous trial (see **Methods Section M2, Step 2**). This decay mechanism ensures that recent contextual associations—whether linked to a US or an extinction engram in the BLA—have a stronger influence on fear responses, while older, unreinforced associations weaken over time. Importantly, the extent of decay of each connection depends solely on the time elapsed since the association was formed, with longer intervals resulting in greater decay, regardless of CS presentations. Although ConFER specifically models cued fear conditioning, the nature of decay of contextual associations suggests a continuous weakening of the fear associated with a context even during periods between trials when no CS is present. By incorporating this decay process, ConFER captures how fear responses dynamically adapt to changing environments.

**Extinction Mechanisms**

When a previously fear-conditioned cue is presented without its associated unconditioned stimulus (US), two key conditions must be met to initiate extinction: the net activation of the Basolateral Amygdala (BLA) is negative, and the US is absent (see algorithm details in **Methods Section M3**). Under these conditions, the model begins the extinction process by checking whether an extinction engram already exists for the given cue. Each cue has a unique extinction engram, defined as a specific set of neuron indices within the positive BLA associated with that cue. If no such engram exists, the model forms one by recruiting a random set of neurons from the positive BLA and storing this set as the extinction engram for that cue. In the model described in the results, extinction engrams consist of four neurons—a value chosen arbitrarily for simplicity, though the model can accommodate varying engram sizes. Once formed, this same set of neurons is consistently activated during any extinction stage for that cue, regardless of extinction context or the number of extinction rounds.

Although the representation of an extinction engram remains stable once formed for a specific cue, the effectiveness of extinction depends entirely on the connection weights between the context representation and the extinction engram. As these connection weights strengthen, the extinction engram becomes more effective at counteracting the negative fear association. However, these context-to-engram connection weights naturally decay over time. Furthermore, each context forms its own independent connection with the extinction engram. When extinction occurs in a new context or after a delay, the model incrementally strengthens the corresponding context-to-engram connection until the cue no longer triggers a fear response.

A key principle in ConFER is that extinction engrams are cue-specific, and they can only be activated when the corresponding conditioned stimulus (CS) is present. The context itself does not activate the extinction engram—it can only build and strengthen its association with the cue-activated engram. Effectively, extinction works as follows: The presence of the cue activates its extinction engram neurons. Once active, these neurons become available for the extinction context to form connections with them. Over repeated exposures, the context strengthens these connections, leading to extinction learning.

If a different CS (say, CS2) is presented in the same extinction context (say, Context A), the extinction engram for the first CS (say, CS1) is no longer active in its absence. Since the CS1 extinction engram is not active, context A can no longer excite the CS1 extinction engram via its learned associations, consequently preventing the activation of downstream projections from the positive BLA to suppress fear. This would lead to a fear response to CS2 when presented in Context A immediately after CS1 was extinguished. CS2 would have to undergo extinction independently by strengthening the association between the extinction context and its own, distinct CS2 extinction engram.

This mechanism explains why extinction in ConFER is both cue-specific and context-dependent. Extinguishing one cue does not automatically generalize to another because each CS is linked to its own extinction engram. While the extinction context may form and strengthen associations with multiple engrams, it cannot activate a specific extinction engram unless the corresponding CS is present.

**Scaling BLA Activations**

ConFER includes a **scaling mechanism** to allow flexibility to adjust the contribution of each BLA engram based on the salience of the associated US. The scaling factors **K1** (for positive US) and **K2** (for negative US) modulate the strength of BLA activation. For example, a footshock would be assigned a higher scaling factor than a puff of air, resulting in a stronger fear response. For the purpose of the demonstration in the results, all USs are set as equally salient.

**Weight Update Mechanism and Final Fear Response Calculation**

ConFER updates connection weights in both the cue–BLA and context–BLA pathways during each trial. The model selectively strengthens connections to the relevant BLA engrams based on the presence of a US. If a US is present, the strength of the connection weights of the cue and context presented in the trial with the neurons representing that specific US in the BLA are separately incremented. The cue weights are incremented more strongly than the context weights in each trial. If a US is absent, ConFER deploys the extinction paradigm, and only increments the weight of the presented context with the extinction engram in the positive BLA. The details of the weight update algorithm are provided in **Methods Section M3.**

The net fear response is computed by summing the scaled activations of the positive and negative BLA populations and passing the result through a non-linear activation function (tanh). This calculation reflects the balance between safety and threat signals in the BLA, producing a final output within a bounded range (−1 to +1), where negative values indicate a fear response and positive values indicate a neutral or appetitive response. Further details are provided in **Methods Section M2.**

**M2. The *execute_training_trial* Function: Stepwise Computational Process in Each Experimental Trial**

The *execute_training_trial* function operationalizes the core mechanisms of ConFER by processing cue-context pairs during each trial. This section outlines the computational steps, providing the detailed algorithm for how the model simulates acquisition, extinction, and fear relapse.

**Step 1: Model Receives Inputs**

Each trial begins with ConFER receiving the following inputs:

1. *cue_index*: Index of the input cue (out of five predefined cues).

2. *context_index*: Index of the input context (out of five predefined contexts).

3. *negative_US_present*: Boolean input indicating whether a negative unconditioned stimulus (US) is present (0 indicates negative US absent, 1 indicates negative US present).

4. *positive_US_present*: Boolean input indicating whether a positive US is present (0 indicates positive US absent, 1 indicates positive US present).

5. *engram_index*: Index of the active BLA US engram (1–4), corresponding to one of the four purple circles in the BLA populations in Figure 1b. This indicates what type of US is present in a given trial, and is set to *None* when no US is present. In the trial demonstrated in figure 1b, this index would be 1, corresponding to the active, yellow filled US engram in the negative BLA representing a shock.

6. *time_from_last_trial*: The time elapsed (in simulated hours) since the last encountered experimental trial.

These inputs define the environmental conditions and stimuli being presented to the model during a given trial.

*Selective Updates to Context and/or Cue Connections with the BLA*

The modify_cue_engram_wts flag in ConFER determines whether cue–BLA connection weights are updated during a trial. By default, the flag is set to FALSE, meaning only context pathway weights are updated when no US is present. When a US is detected, the flag flips to TRUE, allowing both cue and context pathway weights to be updated simultaneously.

- **TRUE**: Cue–BLA and context pathway weights are both updated, strengthening associations with the active BLA engram.
- **FALSE**: Only context pathway weights are updated, reflecting extinction through context–engram associations in the absence of a US.

This selective updating mechanism allows ConFER to dynamically adjust its learning processes based on the presence or absence of unconditioned stimuli.

**Step 2: Temporal Decay of Context–BLA Associations**

To simulate forgetting, ConFER applies an exponential decay function to all existing context–BLA connection weights at the start of each trial, based on the time elapsed since the previous trial. The decay factor, denoted as $\lambda$, is computed from the elapsed time $t$, measured in simulated hours, as shown in Equation 1.

The decay factor weakens contextual associations by proportionally reducing the existing context–BLA connection weights. Specifically, ConFER subtracts a fraction of the current weights, scaled by the decay factor, from the original weights. This process causes contextual associations to decay exponentially with longer delays between trials.

We set the constant $m$=0.001 in Equation 1, resulting in a decay curve that leads to substantial weakening of context weights over approximately 21 days (~500 simulated hours).

$$Decay\ Factor:\ \lambda\ =\ 1\ -\ e^{-mt}\ -\ (1)$$

The decayed context–BLA associations are updated separately for both positive engrams and negative engrams as shown in Equations 2 and 3:

$$Context\text{–}Positive\ BLA_{(15x16)} = Context\text{–}Positive\ BLA\ -\ \lambda * Context\text{–}Positive\ BLA\ -(2)$$
$$Context\text{–}Negative\ BLA_{(15x16)} = Context\text{–}Negative\ BLA\ -\ \lambda * Context\text{–}Negative\ BLA\ -(3)$$

This decay mechanism ensures that recently formed contextual associations retain more influence, while older associations weaken progressively over time.

The decay of contextual associations in ConFER depends solely on time elapsed and is independent of CS presentations during a trial. Within the model, context associations decay exponentially over time, whether or not experimental trials occur during that period. This mechanism assumes that emotional associations with a context weaken continuously with the passage of time, independent of specific cues or experimental manipulations.

If the time elapsed since a context association was formed is sufficiently long (currently set to approximately 21 simulated days), the association will extinguish completely, even in the absence of explicit extinction trials. Experimental trials in the model function primarily as checkpoints to measure the degree of decay that has occurred over the elapsed time, but the decay process itself continues in the background between trials. While the results presented in this paper focus on cued fear conditioning, the ConFER framework can be extended to model contextual fear conditioning. This would require a minor modification to allow trials without explicit cue presentations, while retaining the existing mechanisms for context-based learning and decay.

**Step 3: Computing BLA Activation and Fear Response**

ConFER computes the emotional response to a presented cue–context pair by evaluating the net activation of the positive and negative BLA populations. This net activation is calculated as the weighted sum of the contributions from all neurons in both populations, where each contribution is scaled by the salience of its associated unconditioned stimulus (US).

The scaling factors *K1* and *K2* represent the salience of positive and negative USs, respectively. For simplicity, in the results presented, we assume equal salience for all USs, setting both *K1* and *K2* to 1. The net activation is calculated as shown in Equation 1. In this equation, i denotes the engram index, ranging from 1 to 4 for each BLA population, while j refers to individual neurons within each engram.

$$Net\ BLA\ Activation\ =\ \sum_{i=1}^{4}\sum_{j=1}^{4} K1(i,1) * Positive\ Engram_{i,j}\ -\ \sum_{i=1}^{4}\sum_{j=1}^{4} K2(i,1) * Negative\ Engram_{i,j}\ -\ (4)$$

The result is passed through a tanh function (Equation 5) to constrain the final response to a range of [-1, 1], where -1 represents the maximum fear response and +1 represents the maximum appetitive response. The tanh constant, *n*, is set to 0.01.

$$Fear\ Response\ =\ tanh(\ n\ *\ Net\ BLA\ Activation)\ -\ (5)$$

**Step 4: Check for the Presence of a US, or an Emotional Response in its Absence**

ConFER determines the next steps based on the presence or absence of a US:

*If a US is present (positive or negative):*

1. The model initiates the fear conditioning (if US is negative) or reward conditioning (if US is positive) process by calling the weight_update function (see **Methods Section M3**).

2. The *modify_cue_engram_wts* flag is set to *TRUE*, allowing updates to both cue–BLA and context–BLA connection weights during the trial. This indicates that the presence of a US triggers cue-specific learning, whereas in its default state (*FALSE*), only context–BLA weights are modified, reflecting extinction processes in the absence of a US.

*If no US is present but the Fear Response ≠ 0:*

1. The model initiates the extinction process by evaluating the net fear response to the presented cue–context pair. If the response is less than 0, it indicates a negative (fear) association. If the response is greater than 0, it indicates a positive (appetitive) association. The goal of extinction is to neutralize the association in either case by bringing the net response to 0.

   a. For fear extinction (when the net response is < 0), the model sets *positive_US_present* = TRUE to activate an extinction engram in the positive BLA, counteracting the existing negative association.

   b. For reward extinction (when the net response is > 0), the model sets *negative_US_present* = *TRUE* to activate a reward extinction engram in the negative BLA, counteracting the existing positive association.

2. To ensure that extinction does not reverse the emotional association (e.g., turning fear into reward), the *upper_threshold* is set to the current BLA activation level. The weights of the extinction context with the extinction engram are only updated until the activation of the extinction engram matches this upper threshold. This ensures that extinction reduces the emotional response to 0 without flipping its valence. For example, during fear extinction, the activation of a positive BLA extinction engram via extinction context associations only serves to neutralize the fear activated in th negative BLA by the cue, rather than convert it into a positive (rewarding) association.

3. The *modify_cue_engram_wts* flag remains *FALSE* during extinction, ensuring that only context–engram weights are updated. This reflects ConFER's assumption that extinction relies solely on the context pathway in the absence of a US.

*If neither a US nor an emotional response is present:*

No connection weights are updated, and the trial ends without calling the weight_update function.

**Step 5: Trial Conclusion**

The trial concludes by recomputing the final fear response using the updated connection weights, as described in Step 3.

## M3. The *weight_update* Function: Updating Cue–BLA and Context–BLA connection Weights

The *weight_update* function in ConFER is called once per experimental trial to modify the connection strengths between the cue–BLA and context–BLA pathways. This function governs both acquisition and extinction learning, dynamically adjusting associations between inputs (cues and contexts) and emotional representations in the BLA based on the trial's conditions, and is called within the e*xecute_training_trial* function.

**Function Inputs**

The *weight_update* function takes the following inputs:

1. *cue_index, context_index, negative_US_present, positive_US_present, engram_index*: These are shared inputs from the *execute_training_trial* function that define the cue, context, and the type of US presented during the trial (See **Methods Section M2**, *Step 1*)

2. *cue_to_pos_engram_weights, cue_to_neg_engram_weights*: Matrices representing the connection strengths between cue features and neurons in the positive and negative BLA populations, respectively (See **Methods Section M1**, *Connection Matrices and Pathways*).

3. *context_to_pos_engram_weights, context_to_neg_engram_weights*: Matrices representing the connection strengths between context features and neurons in the positive and negative BLA populations, respectively (See **Methods Section M1**, *Connection Matrices and Pathways*).

4. *net_pos_sum, net_neg_sum*: The net activation levels of the positive and negative BLA populations, computed in the *execute_training_trial* function (See **Methods Section M2,** *Step 3*)

5. *modify_cue_engram_weights*: A flag indicating whether cue pathway weights can be modified during the trial. This flag is set to *TRUE* when a US is present and *FALSE* when no US is present (See **Methods Section M2**, *Step 1*, *Selective Updates to Context and/or Cue Connections with the BLA)*

6. *upper_bound*: The maximum BLA activation level allowed during weight updates, ensuring connection weights with the BLA are not incremented to cause an activation beyond this threshold. This value is set to a high default value, but is adjusted during extinction trials to ensure that extinction engram activation does not exceed the existing emotional activation with the cue in the BLA.

**Identifying Appropriate BLA Engram Neurons for Modifying Association Weights with the Inputs**

The weight update function first identifies the appropriate BLA neurons to activate, via the context pathway and/or the cue pathway, in each trial. The process begins by checking the *engram_index,* which indicates whether an unconditioned stimulus (US) engram is active during a given trial:

- If *engram_index i*s between 1 and 4: This indicates that one of the 4 available USs is present during the trial. This corresponds to an active US engram in either the positive or negative BLA population, based on whether the US is rewarding or fear inducing. The function identifies the four engram neurons associated with this engram_index, referred to as the *selected indices*. For example, in Figure 1b, the yellow-filled circle indicates that engram_index = 1, representing an active engram associated with a shock US present during the trial. The indices of the four neurons within this yellow-filled circle are selected, and the weights between the input cue and context, and these selected neurons will be updated within the *weight_update* function during this fear conditioning trial.

- If *engram_index* is *None*: This indicates that no US is present during the trial, and therefore, no pre-existing US engram is active. If no US is present, and the net BLA activation associated with the input is non-zero, this is an extinction trial (See **Methods M2,** *Step 4*). In this case, the *weight_update* function either forms a new extinction engram if it doesn't already exist for the input cue, or reactivates an existing extinction engram associated with the cue. To form a new fear extinction engram, the model randomly selects four neurons from the entire positive BLA population and stores this set as the extinction engram for that cue. In Figure 1b, the yellow box within the positive BLA population, including 4 random neurons, is an example of an extinction engram for a specific cue. These selected four neurons serve as the *selected indices* for the *weight_update* function, and the weights between the input context and these selected indices will be modified.

**Updating Cue and Context Weights**

The *weight_update* function incrementally adjusts the connection strengths between the input cues and contexts, and BLA neurons during each trial. The updates depend on:

1. Whether an unconditioned stimulus (US) is present, and
2. The activation status of the *modify_cue_engram_weights* flag.

In each trial, ConFER:

- Increments cue–BLA weights by +0.7 if the *modify_cue_engram_weights* flag is set to TRUE, indicating the presence of a US.
- Always increments context–BLA weights by +0.5, irrespective of the flag's state, as long as the net BLA activation remains below the *upper_bound* threshold.

The BLA neurons receiving these updates are the *selected_indices* identified in the previous section, corresponding to the active US engram or extinction engram. The *upper_bound* parameter prevents weight updates from overshooting, ensuring that connection strengths are capped at a reasonable level. By default, this threshold is set to a high value, allowing weight increments to continue across trials. During extinction, however, the *upper_bound* is adjusted to prevent flipping the emotional valence of existing associations that are to be extinguished.

The weight update process across different input conditions is described in the following sections.

*Weight Updates When a US Is Present (Fear and Reward Conditioning)*

When a US is present, the *modify_cue_engram_weights* flag is set to *TRUE,* allowing both cue and context pathways to be updated. The type of US (positive or negative) determines which BLA population and engram are activated.

1. Fear Conditioning (Negative US Present)

- Selected indices: Correspond to the active negative US engram in the negative BLA population.
- Cue weights: Incremented by +0.7 for the selected indices.
- Context weights: Incremented by +0.5 for the selected indices.

- Upper threshold: The updates continue as long as the net negative BLA activation remains below the *upper_bound.*

This process strengthens fear associations, making the cue more likely to trigger a fear response in the future.

2. Reward Conditioning (Positive US Present)

- Selected indices: Correspond to the active positive US engram in the positive BLA population.
- Cue weights: Incremented by +0.7 for the selected indices.
- Context weights: Incremented by +0.5 for the selected indices.
- Upper threshold: The updates continue as long as the net positive BLA activation remains below the *upper_bound.*

This process strengthens reward associations, making the cue more likely to trigger a positive emotional response in the future.

*Weight Updates When No US Is Present (Extinction)*

When no US is present during a trial, the *modify_cue_engram_weights* flag remains *FALSE*, meaning that only the connection strength of the context pathway is updated. The extinction process selectively strengthens the context–engram associations to reduce or neutralize the emotional response evoked by the cue.

3. Fear Extinction (Extinguishing an Existing Negative Association)

- Selected indices: Correspond to the extinction engram in the positive BLA population, which counteracts the negative association.
- Context weights: Incremented by +0.5 for the selected indices.
- Upper threshold: Adjusted to match the magnitude of the net negative BLA activation level, ensuring that extinction neutralizes the fear response without converting it into a positive association.

In this case, extinction reduces the fear response by strengthening the extinction context–extinction engram connection until the fear response reaches zero.

4. Reward Extinction (Positive Association)

- Selected indices: Correspond to the extinction engram in the negative BLA population, which counteracts the positive association.
- Context weights: Incremented by +0.5 for the selected indices.
- Upper threshold: Adjusted to match the magnitude of the net positive BLA activation level, ensuring that extinction neutralizes the reward response without converting it into a fear response.

Similarly, reward extinction reduces the positive emotional response to zero by strengthening the extinction context–extinction engram connection.

**Code Availability:** The code for simulating all findings described in the results is available in the git repository: https://github.com/shreyakr96/CompModelFear

**Acknowledgements:**
The authors would like to thank Dr. Natalie Tronson for her valuable insights on translating findings from animal neuroscience of fear to the learning mechanisms of the model. We are grateful to Dr. Richard Lewis for his constructive comments on the computational model mechanisms and parameters, as well as their impact on model simulations. We also appreciate Dr. Elizabeth Duval's thoughtful contributions regarding the clinical implications of the model and its connections to anxiety and stress-related disorders.

**Author Contributions:**
S.K.R. reviewed animal neuroscience findings, conceptualized the neurally constrained modeling framework, developed the computational code to simulate empirical phenomena and generate novel hypotheses, and wrote the manuscript. T.A.P. supervised the project, helped motivate and refine the connectionist architecture of the model, provided feedback on model assumptions, parameters, and simulations, and contributed to the writing and editing of the manuscript.

**Competing Interests:**
The authors declare no competing interests.